# SDN Enabled and OpenFlow Compatible Network Performance Monitoring System

Sayyaf Haider Warraich[1], Zeeshan Aziz[1], Dr. Hasnat Khurshid[1], Rashid Hameed[2], Abdul Saboor[3]
Muhammad Awais[2]

[1]National University of Sciences and Technology, Islamabad, Pakistan

[2]School of Electrical Engineering and Computer Science, National University of Sciences and Technology (NUST), Islamabad, Pakistan

[3] Thomas Johann Seebeck Department of Electronics, Tallinn University of Technology, Tallinn, Estonia
Email:[sayyafhaider.msee22, zeeshanaziz.msee22]@students.mcs.edu.pk, hasnat@mcs.edu.pk, [rashid.hameed,
mawais.msee17seecs]@seecs.edu.pk, abdul.saboor@taltech.ee

*Abstract*—Network performance monitoring holds a pivotal role in improving the overall network performance. It is essential to monitor the traffic statistics at Internet eXchange Points (IXPs) to optimize the traffic flows. The existing monitoring system either lacks usability or programmability. We present a Software Defined Networking (SDN) enabled and OpenFlow (OF) compatible network performance monitoring system (SDX-Manager) to improve the usability or programmability of IXP simultaneously. The proposed system integrates traditional IXP-Manager with the SDN controller and Grafana to better visualize the traffic. It provides a dashboard to the network administrator for defining rules over IXP, such as port blocking and traffic redirection. Furthermore, it facilitates the administrative tasks that include adding, modifying, and deleting the users and IXP objects (Ports, Switches). The proposed SDX-Manager is deployed and tested in a virtual and physical environment using OF enabled Allied Telesis Switch. Results indicate that SDX-Manager has improved network analysis using different statistics such as packets in/out per second, bits in/out per second, scrape duration, and query count.

Keywords—SDX-Manager, Internet eXchange Point (IXP), Software Defined Networking (SDN), Programmability

## I. INTRODUCTION

The internet is expanding at a faster pace in the current era, with a growing number of users and devices connected to it [1]. It is used in daily communication, education, gaming, industrial automation, and self-driven automation [2]. Wireless communications facilitate tale-health, defense, emergency, sports, and fitness applications [3], [4]. This massive expansion of the internet brings various challenges for a network administrator such as network management, network configuration, network performance, and security [5]. In addition to that, there are also many customer-oriented problems: internet speed, bandwidth, latency, and connectivity. [6]. One of the major causes of the above problems is the network congestion in the presence of high-quality media content (high-quality graphics and high definition videos) shared over the internet. Moreover, the presence of resources hungry applications such as online gaming and remote robotic surgery over the traditional network also creates a bottleneck. The bandwidth optimization techniques help in avoiding the network congestion that eventually resolves the customer-oriented problems [7].

Internet eXchange Point (IXP) is an efficient solution to optimize network bandwidth. It provides a platform to Inter-net Service Providers (ISPs) and Content Delivery Network (CDN) providers to share the traffic via a common point. It reduces the number of connections among different ISPs, thus, reduces the overall complexity of the network. Also, it facilitates the internal routing of local traffic without going over the international line. As a result, it decreases the number of connections, data routing to specific links, latency, and transit cost due to international routing. Along with band-width optimization, IXP helps in improving the security of the network as well. Generally, the routing of local traffic internationally enhances the risk of compromising the security of the data. IXP eliminates such risks by routing the traffic to the designated local ISPs and CDNs. [8]

The IXPs have resolved many problems that were present in a network. However, there are particular challenges associated with the existing IXPs. For example, the existing IXPs are unable to support a wide range of data plane policies such as application-specific peering, redirection of traffic to middleboxes, inbound traffic engineering, and wide-area server load balancing. Similarly, the existing IXPs use Border Gateway Protocol (BGP) for inter-domain routing, which offers limitations like routing limited to destination IP prefix and indirect path selection mechanisms. Software Defined Networking (SDN) was introduced in the IXPs to overcome the limitations posed by the traditional IXPs [9]. The SDN enabled IXPs challenges are addressed by intro-ducing centralized programmability on the Open Flow (OF) controller in inter-domain routing [10]. However, it is unable to provide crisp statistics details to the system administrator. Furthermore, it requires extensive training to control the SDN enabled system that is still prone to human errors. All of such factors reduce the overall system usability. In order to address the above mentioned challenges, we provide the following contributions in this paper:

We have developed a Laravel based SDN enabled IXP-manager (SDX-Manager) that adds the programmability over the traditional IXP-Manager.

We have integrated a more delicate traffic statistics visu-alizer for analysis.

We have developed a full-stack Graphical User Interface (GUI) dashboard, contrarily to the traditional Command Line Interface (CLI), to define routing and peering rules. It results in increasing overall system usability.

We have added a few Access Control Lists (ACLs) for applications like port blocking, port mirroring, and traffic redirection. Furthermore, a network manager can create custom ACLs on top of SDX-Manager.

We have tested our system on virtual as well as on physical topologies.

The rest of the paper is organized, as Section II presents the background and motivation of work. Section III explained the overall design and implementation of SDX-Manager. The results of the proposed system are elaborated in Section IV. Finally, Section V concludes the work with future directions.

## II. BACKGROUND AND MOTIVATION

This paper aims to improve the usability and programma-bility of the IXP. For the reader's understanding, we provide an overview of IXP, IXP-Manager, and SDN in this section.

### A. IXP

The internet comprises of thousands of individual networks interconnected with each other and sharing traffic. The IXP provides a physical platform where different networks connect to exchange their traffic. In an IXP, the first step is to estab-lish peering with other networks. The peering is established between two parties based on mutual agreement of both parties that may include sharing their respective details as well. As a result, it resolves the hassle of setting up a connection every time for traffic sharing with the rest of the networks. Furthermore, this process reduces the cost of transit point and network operations, improves system efficiency with better control, reduce latency by direct connections, and provide security of the data by keeping traffic in a local network [6]. An operator or administrator manages IXP operations. The general responsibilities of an operator includes ensuring the server availability, IXP management and maintenance of the infrastructure without interrupting the ISP's traffic. IXP operation team's primary responsibility is to maintain the physical infrastructure of IXP, such as cabling, website, servers, and switching equipment. They also provide technical assistance on the issues regarding IXP to ISPs. IXP is installed at a neutral point to gain maximum benefits of the service. One of the major concept of the IXP is to get rid of the monopoly of different service providers and deliver multiple utilities to ISPs. It comprises of layer2/layer3 switches [11]. Currently, 956 IXPs are installed in 159 countries as shown in Figure 1 [12].

Traditional IXP architecture consists of at least one switch to establish a layer-2 switching fabric. At fabric, multiple ISPs and local content delivery providers share their local traffic with IXP members, according to the peering rules defined by IXP. Figure 2 illustrates the traditional IXP architecture in which four different ISPs share traffic locally. All the ISPs are peered with each other using a traditional switch and formed an IXP switching fabric one. Similarly, three other ISPs peered with each other using another switch to form an IXP switching fabric two. These different switching fabrics further connected through a link to exchange their local data [13].

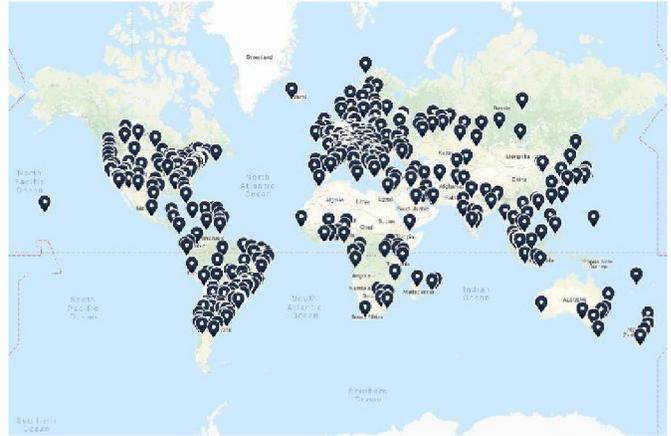

Fig. 1. IXPs in the world.

Traditional IXP has made traffic sharing between different ISPs more convenient. However, its management is a chal-lenging task and requires extensive training. Besides that, the overall control of an IXP via traditional CLI is prone to human errors. All of such challenges result in degrading the usability of the system. IXP-Manager resolves such problems by managing the operations of IXPs [14].

### B. IXP-Manager

The IXP-Manager is a full-stack web application that fa-cilitates the traditional IXP system management by providing a GUI. Due to the usability of IXP-Manager, it is currently deployed in 93 IXPs around the world [14]. IXP-Manager provides a full solution for the IXPs administration by pro-viding a separate admin and customer portal. An admin can create different roles using this portal. Additionally, it helps in visualizing the traffic statistics. The IXP-Manager has undoubtedly improved the usability and adaptability of the system. However, it lacks the programmability on the fabric, which can be achieved through the addition of SDN enabled controller in the IXP-Manager.

### C. SDN

SDN architecture improves the network controllability by separating the control plane from the data plane. Hence, it provides abstraction of lower-level functionality [15] that helps network administrators to (programmatically) initialize, mon-itor, control, and manage network behavior via OF protocol. One of the significant advantages of separating the data and control plane is to get rid of vendor-specific equipment. It also provides a central management system to control all the switches [16]. A general SDN architecture is presented in Figure 3 that consists of the following parts:

Application Plane

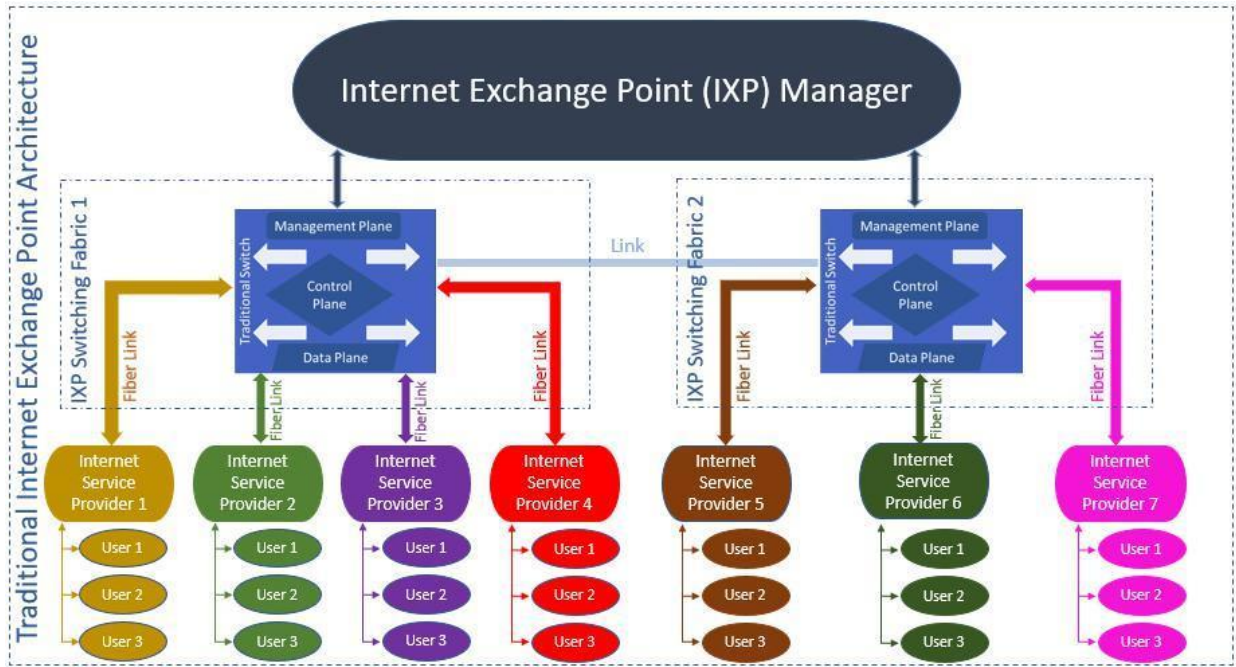

Fig. 2. Traditional IXP architecture.

Control Plane
Data Plane
OF Protocol
North-Bound Application Program Interfaces
(APIs) South-Bound APIs

The top layer of SDN architecture consists of multiple net-work applications. For example, the applications can manage the network or monitor the statistics for decision making. These layers communicate with the SDN controller via APIs. The control plane is an instruction set or decision entity, which runs on the top of the OF switch. SDN Controller is installed in this layer for network management to control the packet flows. The Data plane includes forwarding devices where the instructions are taken from the controller using southbound APIs. These devices forward data using the control instructions.[17].

SDN makes the IXP programmable by introducing the application layer on top of the controller. However, the SDN controller without programmability is not that easy to use for an operator's perspective. A network operator must require the programmability as well as usability to manage the network.To the best of our knowledge, the TouSix manager is the only effort done in this direction [18]. However, it works for a pre-defined topology that limits its adaptability. Therefore, to simultaneously address the usability and programmability of IXP, we have developed a hybrid solution (SDX-Manager) by integrating the SDN controller with the IXP-Manager. In contrast to the TouSix manager, the SDX-Manager can handle dynamic topologies. The details of SDX-Manager is presented in the next section.

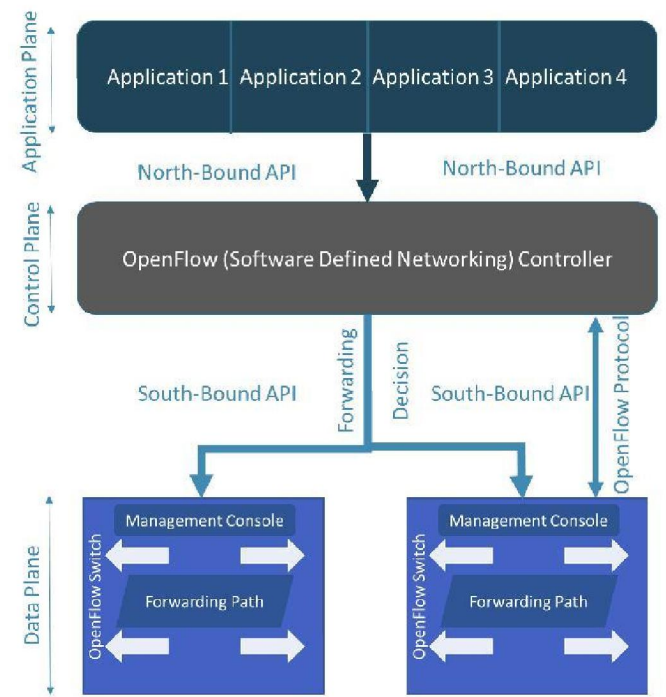

Fig. 3. SDN architecture.

## III. SDX-MANAGER DESIGN

SDX-Manager uses the SDN paradigm in the domain of networking to enhance the programmability of IXP. In SDX-Manager, an SDN controller support is added to tra-ditional IXP-Manager. Moreover, the IXP-Manager has been redesigned to define rules from the top of the controller. It results in optimizing the traffic paths, preventing the security attacks using ACLs, and establishing a BGP peering of ISPs. Figure 4 outlines the architecture of the proposed SDX-Manager. The graphical dashboard of SDX-Manager provides the following benefits:

It provides the ability to define the rules through the SDN controller to control the traffic.

It provides the provision to add new Autonomous System (AS) and establish its peering with other ASs.

It provides the flexibility to create, modify, and delete different users such as admin, moderator, a regular user (customer), and their roles.

It provides the facility to monitor the live statistics of traffic, users, and other inventories such as racks, switches.

SDN controller live status provides the status of the controller along with listening ports. Additionally, it graphs the Central Processing Unit (CPU) resource utilization and memory resource utilization. For this task, we have written a PHP script based on the phpseclib library [19]. In the third part, the administrator can add/select Virtual Local Area Network (VLAN) and OF switch. VLAN establishes communication between peered ASs, whereas OF switch provides a switching mechanism. The switch interface addition helps in establishing the peering between ISPs on top of the OF switch with the SDN controller. Hence, the controller generates the rules on top of switches to provide security by blocking malicious traffic on the runtime. Furthermore, SDX-Manager offers static load balancing, i.e., dividing the load between ports by redirecting the load of the congested port. It can also be used to mirror traffic for running data processing algorithms, such as machine learning or artificial intelligence. The SDX-Manager pushes different rules to Faucet controller by using YAML Ain't Markup Language (YAML) files. In the proposed scheme, we have used the faucet controller because of its compact size, high availability, scalability, and independence from an external database for connectivity [20]. Furthermore, it offers the provision of multiple tables, containing a smaller number of flows. The overall SDX-Manager setup consists of the following steps.

### A. Setup and configuration

The setup and configuration process consists of the follow-ing parameters.

Kernal Virtual Machine (Virtual-Box) Faucet controller
Grafana server
Prometheus data
base IXP-Manager
Quagga software routing suite
Open Virtual Switch (OVS)
IPERF network performance measurement

KVM is used to set up the virtual environment for SDX-Manager in which the Faucet controller manages the control functionality. Grafana server is a visualization tool to plot graphs that links with the Prometheus database, which stores statistical data for graph processing. IXP-Manager is used as a tool to manage the IXPs network. Quagga is a virtual router that is combined with a virtual OF compliant virtual switch for testing. Lastly, IPERF tests the bandwidth while using dummy traffic in our scenario. The overall setup is created in a virtual environment with the configuration of the faucet controller, gauge controller, Prometheus, and Grafana over the IXP-Manager. For the configuration of the controller, we have created a new table in the existing schema of the IXP-Manager database using the Laravel framework. It leads to the addition of a new tab (Faucet Controller) on the dashboard. Additionally, a second tab named live status is added below the faucet controller tab to fetch the running status of the faucet SDN controller. It shows the listening port status, memory utilization, CPU utilization, and running status of the faucet and gauge processes.

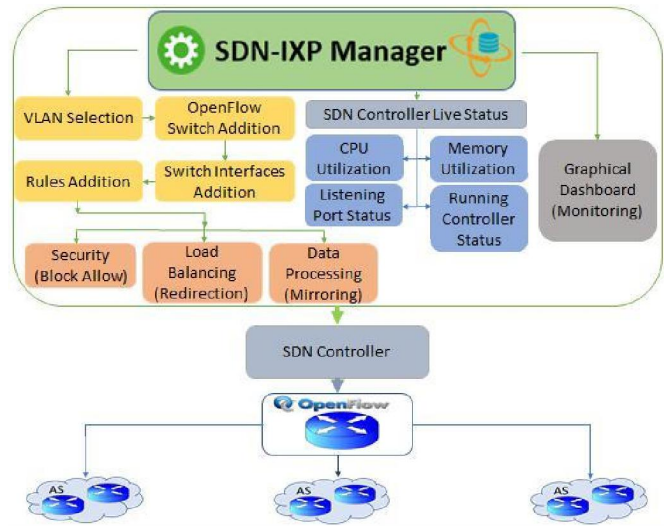

Fig. 4. SDX-Manager architecture.

### B. Traffic statistics

The major outline of this research is monitoring and analysis of network traffic such as data inflow and outflow from each port of the switch, packet drop rate, errors per second, and controller information (CPU usage, memory, cold restart). For that purpose, a separate statistical collector is required to integrate with the traditional IXP-Manager. The idea is to redirect the traffic report towards the configured collectors for better visualization. Therefore, the SDX-Manager integrates the Grafana dashboard. The reason for using the Grafana is due to its ability to design custom and interactive dashboards for stats monitoring and analysis. The admin or customer can navigate to the statistical portal to monitor the traffic and controller information by clicking the recently added Grafana button in SDX-Manager.

### C. Peering and rules generation

One of the major objectives of an IXP administrator is to establish peering between the ISPs by defining the peering rules. Traditional IXPs use vendor-specific equipment that is unable to support a wide range of data plane policies. Hence, it limits the flexibility to define peering rules based on traffic statistics such as in the case of traffic overload on a certain port. SDX-Manager overcomes such deficiencies by separating the control plane from the data plane using OF enabled controller. In the proposed SDX-Manager, an administrator defines the peering instructions using the GUI of SDX-Manager. To facilitate the communication between the application and controller, SDX-Manager generates a YAML file containing the rule and peering information. Faucet controller analyzes the YAML and implements the desired actions such as peering, load balancing. A YAML file requires the information of VLANs, switches, interfaces, ACL and peering details as shown in the Fig 5. For that, an administrator can select the existing VLAN, switch, and interface or create a new entry of one or more of these by initializing parameters shown in Table I.

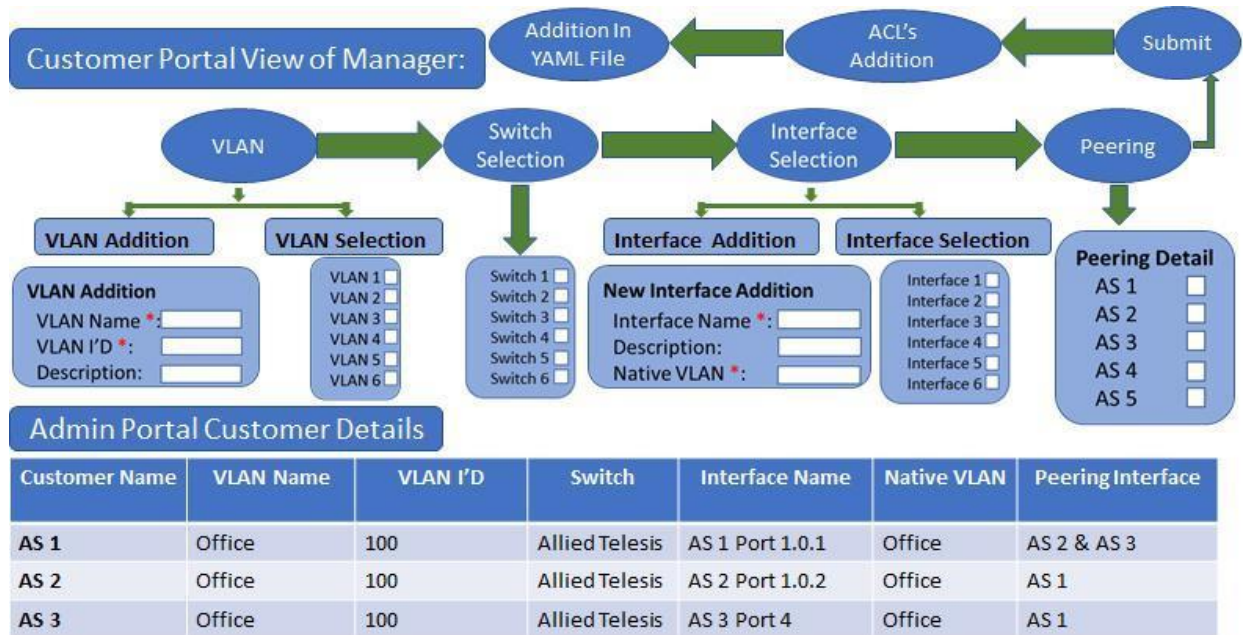

Fig. 5. Flow diagram of customer portal in SDN enabled IXP-Manager.

TABLE I
YAML FILE PARAMETERS FOR PEERING

| Name | Parameters |
|---|---|
| VLAN | VLAN Name<br>VLAN ID<br>VLAN Description |
| Switch | OF Switch Name<br>OF Datapath ID (dpn id).<br>OF Hardware Name |
| Interface | Interface Name<br>Interface Description<br>Native VLAN |
| ACL | Block Ports<br>Mirror Ports<br>Redirection<br>Allow All |

As defined earlier, The rules generation in a network is the most important task for a network engineer, as these rules are required to prevent a security breach, multiple cyber-attacks, and redirect the data based on congestion. For the facilitation, we have embedded a few widely used rules in

SDX-Manager via ACLs, as shown in Table I. Block port results in clogging the port carrying the malicious data. In this way, it prevents the Distributed Denial of Service (DDoS) attacks. Mirror port creates an identical port to analyze the traffic while allow-all ACL floods the packets to all ports. The redirection allows redirecting traffic based on defined VLAN, switch, and interface. It is generally used for load balancing as well as optimal routing. A YAML file generated via SDX-Manager for traffic mirroring is presented in Fig. 7. In addition to predefined ACLs, the SDX-Manager provides the flexibility to create custom ACLs on the platform. The analysis of the traffic is presented in the next section.

## IV. RESULTS

SDX-Manager primarily aims to provide network statistics for the analysis. Therefore, we have integrated Prometheus and Grafana with the SDX-Manager to visualize the minor network traffic details. Fig. 6 provides the statistics of memory profile, scrap duration, head chunks, etc. The following are the details of Fig. 6:

a) This figure provides the samples appended per second, which is the speed rate of the head sample appended to the time-series database. This scenario shows a constant (62) samples appended in the time frame of 30 seconds.

b) This figure provides the scrap duration of the Faucet and Gauge controller. Scrap duration is the loading time of the YAML file and instances. During this analysis, the scrape duration is, on average less than 25 ms.

c) This figure provides the memory profile to show the virtual and resident memory in one window. It shows resident memory a maximum of up to 85.473 MebiByte (MiB) and virtual memory 291 MiB when the controller is running on the server.

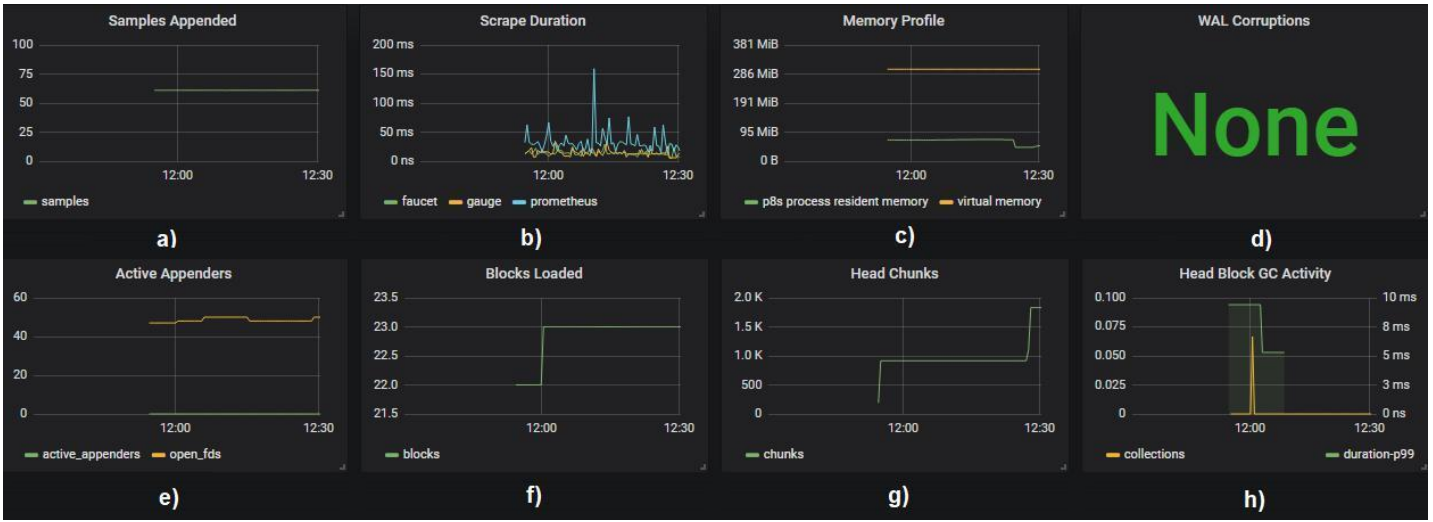

Fig. 6. Network statistics using SDX-Manager.

```
vlans:
    office:
        vid: 100
        description: "Research network"
dps:
    sw1:
        dp_id: 0x1
        hardware: "Open vSwitch"
        interfaces:
            1:
                name: "AS1"
                description: "port 1.0.1"
                native_vlan: office
            2:
                name: "AS2"
                description: "port 1.0.2"
                native_vlan: office
                acls_in: [mirror, allow-all]
            3:
                name: "AS3"
                description: "port 4"
                native_vlan: office
            4:
                name: "AS4"
                description: "port 5"
                native_vlan: office
acls:
    mirror:
        - rule:
            dl_type: 0x800      # IPv4
            ip_proto: 1         # ICMP
            actions:
                allow: False
                mirror: 4
        - rule:
            dl_type: 0x86dd     # IPv6
            ip_proto: 58        # ICMPv6
            actions:
                allow: False
                mirror: 4
    allow-all:
        - rule:
            actions:
                allow: True
.
```

Fig. 7. A YAML file generated via SDX-Manager for port mirroring.

d) This figure provides the Write-Ahead Log (WAL) graph to check the status of the WAL segment. The WAL generates a recovery mechanism in case of node or cluster failure. In this scenario, there is no WAL corruption due to failures.

e) This figure provides an active appender to monitor the number of active appender transactions. Generally, the appender instructions write the event data to target des-tinations.

f) This figure provides the block loaded graph, which ana-lyzes the number of blocks loaded in the database.

g) This figure provides the chunk's information. Prometheus uses chunks to store the data in the form of time series.

h) This figure provides the head block GC activity. It plots the head block collection against the time.

Some additional traffic statistics are presented in Fig. 8. Fig. 8a visualizes the compaction process that merges multiple blocks (in the database) into one block to avoid bulk scan. The reload count graph in Fig 8b shows how periodically database is reloaded during the configuration process. The duration of different queries is plotted in Fig. 8c. Similarly, rule group evaluation and activity are presented in Fig. 8d and Fig. 8e. Finally, Fig. 9a and Fig. 9b plot the system parameters such as CPU usage and used file descriptors. It justifies that the Faucet controller consumes fewer resources, such as a maximum of 0.1091% in this case. In addition to these statistics, Grafana offers the flexibility to design the custom dashboards to visualize more/different traffic parameters.

We have tested our proposed SDX-Manager in the virtual as well as in the physical environment. First, we have an-alyzed the results in the form of two-nodes topology. After that, the same topology is created and tested in a physical environment using the Allied Telesis switch. Finally, we adopt a bigger topology to examine the adaptability, scalability, and functionality of the SDX-Manager.

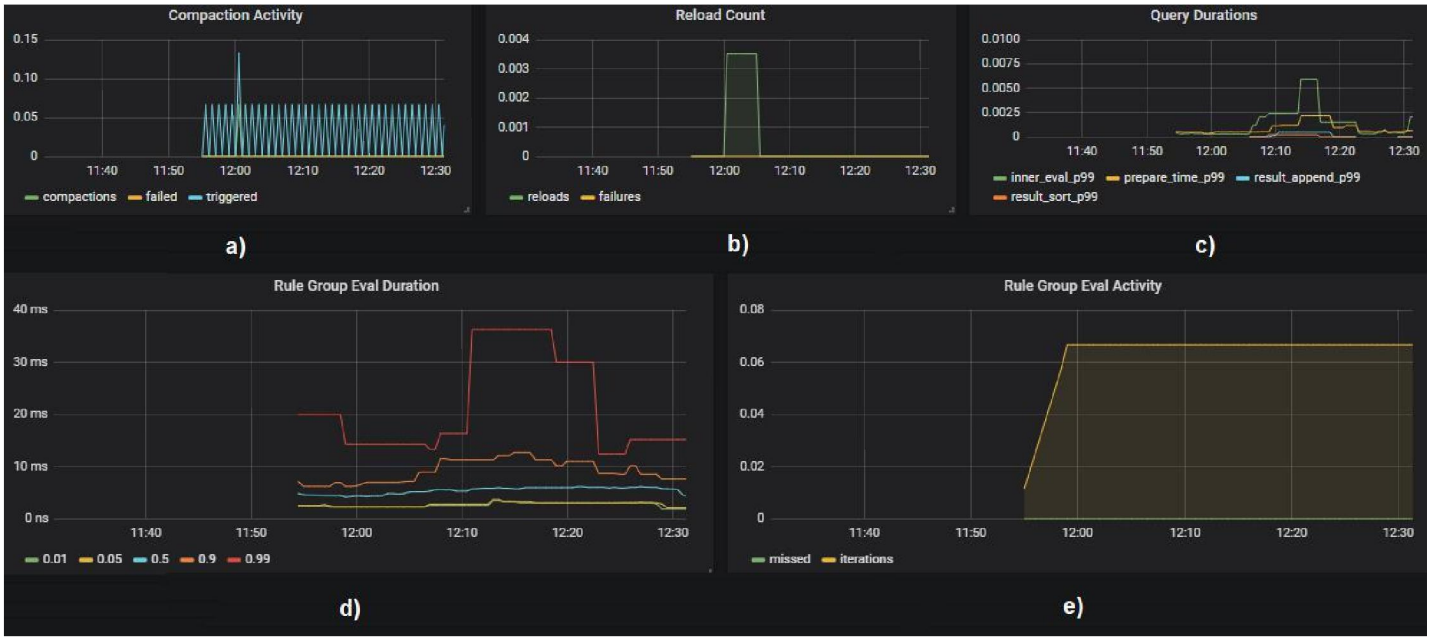

Fig. 8. Additional network statistics using SDX-Manager

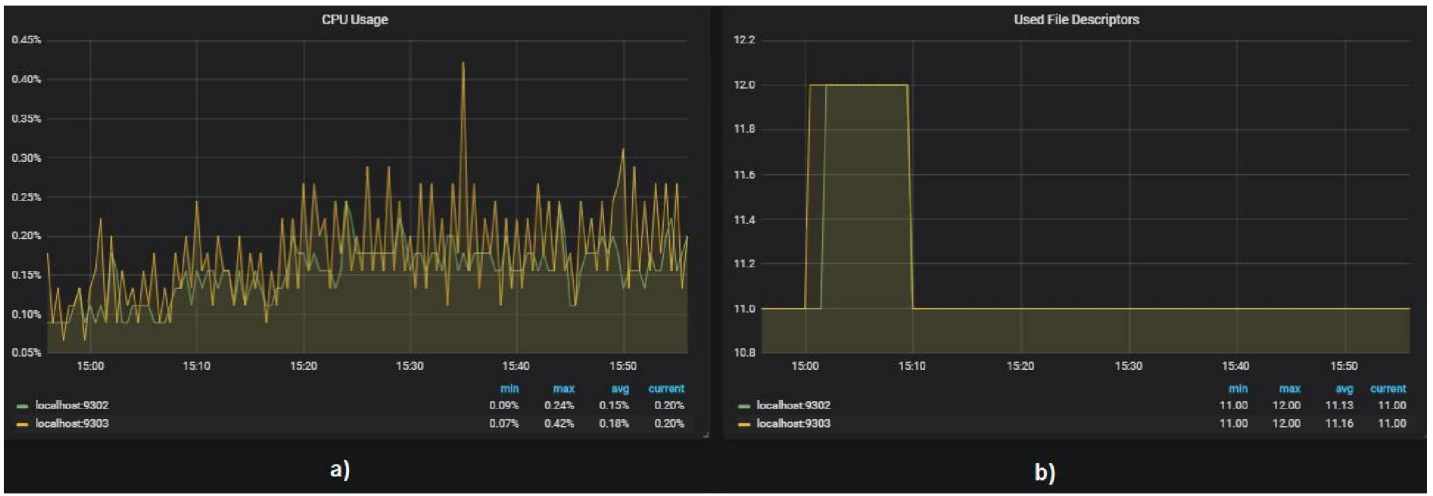

Fig. 9. System statistics using SDX-Manager

## A. Virtual topology using two nodes

In this topology, we created two separate ASs (AS-1, AS-2) in the virtual environment, as shown in Fig. 10. The connection between two ASs is created using a distributed virtual switch named OVS. The OVS further connects with the SDN controller using SDX-Manager. It helps in configuring the ASs and establish the BGP peering for communication. The sole purpose of the creation of two virtual servers is to test the SDX-Manager. One of the desired requirements of an ISP is to visualize the port in/out status. Therefore, Fig. 11 shows the bits in/out per second while transferring the data from AS-2 to AS-1. The system is supporting the throughput of more than 1Gbps. Similarly, Fig. 12 display the packets in/out per second during the transmission. It is observed from the statistics that more than 75k packets per second are received at AS-1.

## B. Physical topology using two nodes

We created the same topology, as in the previous section, in the lab environment. In this scenario, we connect two primary services provides in Pakistan (Nayatel as AS 2 and PTCL as AS 1) with Allied-Telesis switch, as shown in Fig. 13. For the evaluation of virtual topology, we examine the same results of Bits in/out and Packets in/out in this topology. Figures 14 and 15 displays the same traffic trends, as in the similar virtual topology. It shows that SDX-Manager works seamlessly from topology nature.

## C. Virtual topology using four nodes

In this section, we aim to check the scalability support of SDX-Manager using four ASs in a virtual environment. All the ASs are connected via OVS, as shown in Fig. 16. The

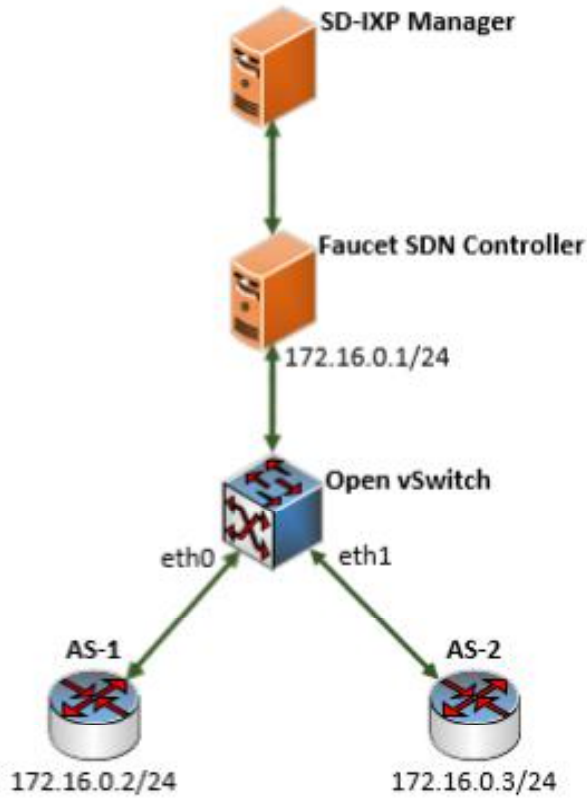

Fig. 10. Virtual topology using two nodes.

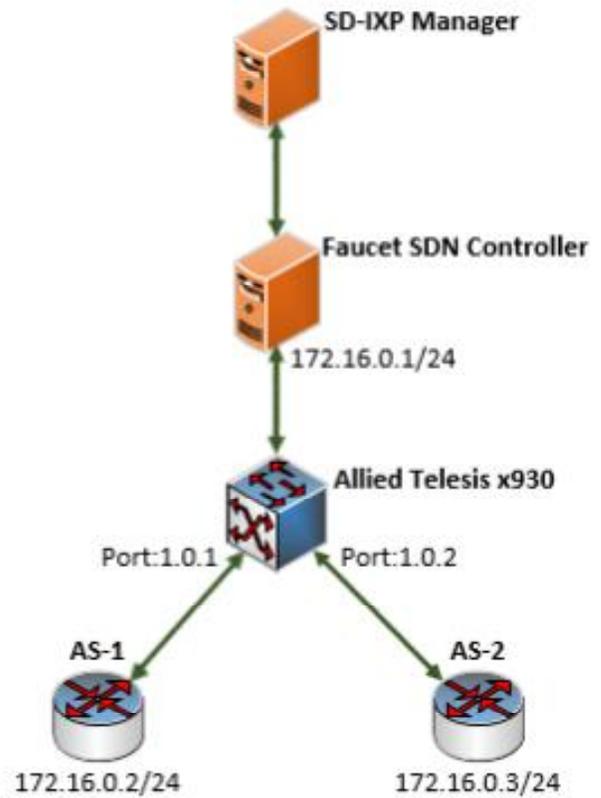

Fig. 13. Physical topology using two nodes.

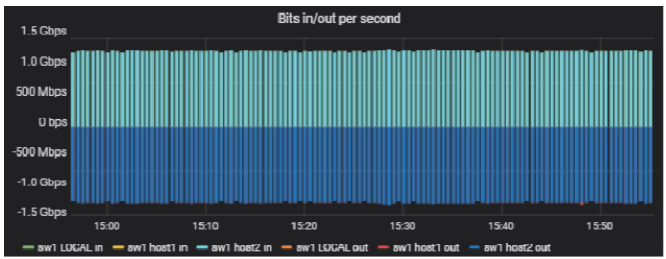

Fig. 11. Bits in/out in a virtual topology of two nodes.

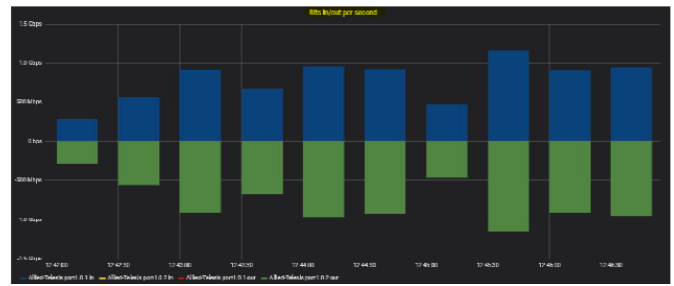

Fig. 14. Bits in/out in a physical topology of two nodes.

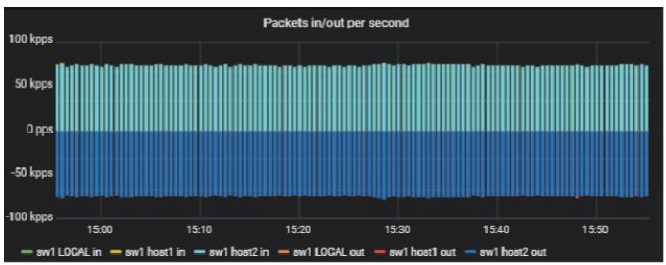

Fig. 12. Packets in/out in a virtual topology of two nodes.

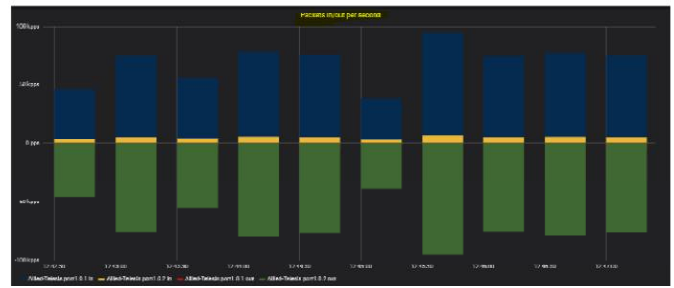

Fig. 15. Packets in/out in a physical topology of two nodes.

statistics of bits and packets are graphed via SDX-Manager in Fig. 17 and Fig. 18, respectively. The proposed system works well in this virtual environment, which envisions its possibility to work in large physical topologies. Currently, we have only tested SDX-Manager in two ASs physical topology due to the limitation of resources. However, we are in talks with Pakistan

Telecommunication Authority (PTA) to test SDX-Manager on the IXP.

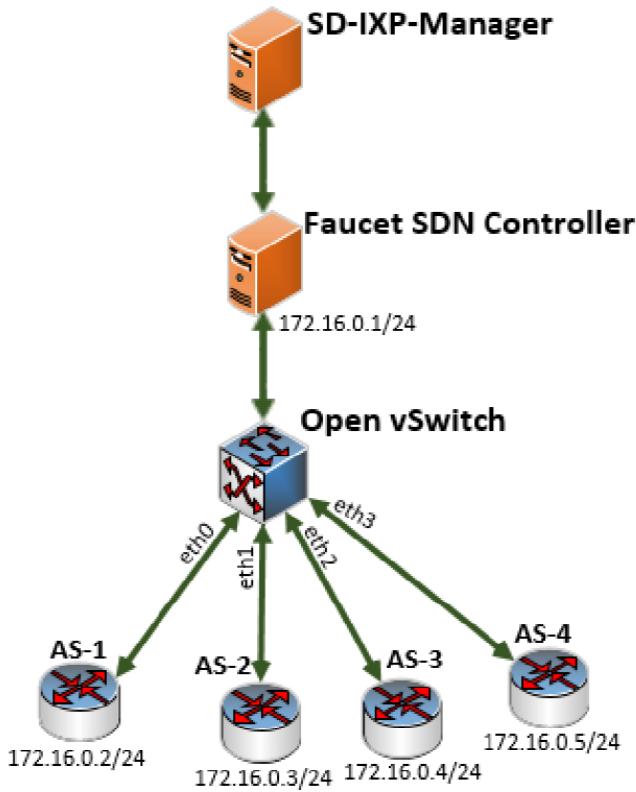

Fig. 16. Virtual topology using four nodes.

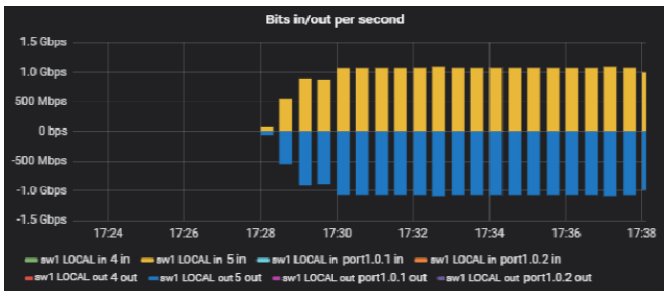

Fig. 17. Bits in/out in a virtual topology of two nodes.

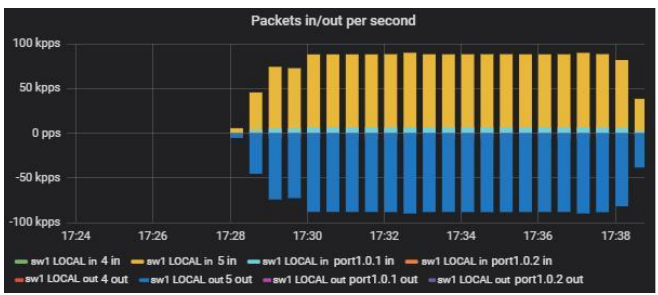

Fig. 18. Packets in/out in a virtual topology of two nodes.

## V. CONCLUSION AND FUTURE DIRECTIONS

This paper presents an SDX-Manager to control and monitor IXP peering. It provides a user-friendly GUI (dashboard) to the network administrator that helps in defining the network rules using an SDN controller. Besides, it provides the provision of the SDN controller to establish effective communication between different ISPs dynamically. The proposed solution is tested in virtual and physical environments. Results show that the SDX-Manager is capable of providing crisp traffic statistics to analyze traffic. Furthermore, the network administrator can use various pre-defined and custom-built ACLs for applications like port mirroring, port blocking, traffic redirection, etc.

Currently, the SDX-Manager only supports the Faucet controller due to its lightweight size. However, a rapid increase in technology motivates multiple researchers, vendors, and organizations to develop controllers. These controllers will come up with functionalities to cater to the networking demands in a better way. Therefore, our future plan is to add the support of multiple/different controllers in SDX-Manager. Secondly, multiple versions of the OF protocol have been released with an advanced set of standards. Each version brings new rules for adequate control. Therefore, we aim to integrate these rules while redesigning the graphical interface.

## ACKNOWLEDGMENT

In the end, the authors would like to express their gratitude to Bitsym Private Limited for allowing the authors to use the facility and Allied Telesis Switch for validating the test results. Moreover, the authors would also like to thanks Dr. Marc Bruyere, Post Doctorate University of Tokyo for extending his technical support during this research work.